\documentclass{aa}
\usepackage[obeyspaces]{url}
\usepackage{hangcaption} 
\usepackage{natbib} 
\usepackage{graphicx}

\begin{document}
\title{Pulsar searches and timing with the SKA} 
\author{R.~Smits \inst{1} \and M.~Kramer \inst{1} \and B.~Stappers
  \inst{1} \and D.R.~Lorimer \inst{2,3} \and J.~Cordes \inst{4} \and
  A.~Faulkner \inst{1}} 
\offprints{R. Smits \email{Roy.Smits@\allowbreak
  manchester.\allowbreak ac.uk}} 
\institute{Jodrell Bank Centre for Astrophysics, University of
  Manchester \and Department of Physics, 210 Hodges Hall,
  West Virginia University,
  Morgantown, WV 26506, USA \and National Radio Astronomy Observatory, Green Bank \and
  Astronomy Department, Cornell University, Ithaca, NY}
\date{Received «date» / Accepted «date»} 
\abstract { 
The Square Kilometre Array (SKA) is a planned multi purpose radio
telescope with a collecting area approaching 1 million square metres.
One of the key science objectives of the SKA is to provide exquisite
strong-field tests of gravitational physics by finding and timing
pulsars in extreme binary systems such as a pulsar-black hole
binary. To find out how three preliminary SKA configurations will
affect a pulsar survey, we have simulated SKA pulsar surveys for each
configuration. 
We estimate that the total number of normal pulsars the SKA will
detect, using only the 1-km core and 30 minutes integration time, is
around 14\,000 normal pulsar and 6\,000 millisecond pulsars.  We
describe a simple strategy for follow-up timing observations and find
that, depending on the configuration, it would take 1--6 days to
obtain a single timing point for 14\,000 pulsars. Obtaining a single
timing point for the high-precision timing projects of the SKA, will
take less than 14 hours, 2 days, or 3 days, depending on the
configuration. The presence of aperture arrays will be of great
benefit here.  We also study the computational requirements for beam
forming and data analysis for a pulsar survey.  Beam forming of the
full field of view of the single-pixel feed 15-m dishes using the 1-km
core of the SKA requires about 2.2$\times$10$^{15}$ operations per
second.  The corresponding data rate from such a pulsar survey is
about 4.7$\times$10$^{11}$ bytes per second. The required
computational power for a deep real time analysis is estimated to be
1.2$\times10^{16}$ operations per second. For an aperture array or
dishes equipped with phased array feeds, the survey can be performed
faster, but the computational requirements and data rates will go up.
}
\authorrunning{Smits et al.} 
\titlerunning{Draft: Pulsar searches and timing with the SKA}
\maketitle

\section{Introduction}
There are, arguably, no other astronomical objects whose discovery and
subsequent studies provides more insight in such a rich variety of
physics and astrophysics than radio pulsars. Pulsars find their
applications \citep[see e.g.][]{lk05} in the study of the Milky Way,
globular clusters, the evolution and collapse of massive stars, the
formation and evolution of binary systems, the properties of
super-dense matter, extreme plasma physics, tests of theories of
gravity, the detection of gravitational waves, and astrometry, to name
only a few areas. Indeed, many of these areas of fundamental physics
can be best -- or even only -- studied, using radio observations of
pulsars.

To harvest the copious amount of information and science accessible
with pulsars, two different types of observations are
required. Firstly, suitable pulsars need to be discovered via radio
surveys that sample the sky with high time and frequency
resolution. Depending on the particular region of sky to be covered
(e.g.~along the Galactic plane vs.~higher Galactic latitudes),
different technical requirements may be needed. Secondly, after the
discovery, a much larger amount of observing is required to extract
most of the science using pulsar timing observations, i.e.~the regular
high-precision monitoring of the pulse times-of-arrival and the pulse
properties. Again, depending on the type of sources to be monitored
(e.g.~young pulsars in supernova remnants vs.~millisecond pulsars in
Globular clusters) the requirements are different.

In the past 40 years, astronomers have impressively demonstrated the
potential of pulsar search and timing observations
\citep[e.g.][]{ht75}. However, the next 10-15 years promise to
revolutionise pulsar astrophysics in a way that is without parallel in
the history of pulsars or radio astronomy in general. This revolution
will be provided by the next generation radio telescope known as the
``Square-Kilometre-Array'' (SKA) \citep[e.g.][]{Terzian2006}. The SKA
will be the largest telescope ever built, with a maximum
baseline of 3\,000+\,km and about a factor of 10--100 more powerful
(both in sensitivity and survey speed) than any other radio telescope.
The SKA will be particularly useful for pulsars and their
applications in astrophysics and fundamental physics. Due to the large
sensitivity of the SKA, not only a Galactic census of pulsars can be
performed, but the discovered pulsars can also be timed more precisely
than before. As a result, the pulsar science enabled by the SKA and
described by \citet{Cordes2004} and \citet{Kramer2004} includes:
\begin{itemize}
\item Finding rare objects that provide the greatest opportunities as physics
  laboratories. Examples include:
 \begin{itemize}
 \item Binary pulsars with black hole companions, which enable strong-field
   tests of General Relativity.
 \item Pulsars in the Galactic Centre, which will probe conditions near a
   $3\times10^6\,\mathrm{M}_\odot$ black hole.
 \item Millisecond pulsars that can act in an ensemble (known as
   Pulsar Timing Array) as detectors of low-frequency (nHz)
   gravitational waves of astrophysical and cosmological origin
 \item Millisecond pulsars spinning faster than 1.4\,ms that can
   constrain the equation-of-state of super-dense matter.
 \item Pulsars with translational speeds in excess of
   $10^3$\,km\,s$^{-1}$ that can probe both core-collapse physics and
   the gravitational potential of our Galaxy.
 \item Rotating radio transients \citep[see][]{mll+06}
 \item Intermittent pulsars \citep[see][]{klo+06} 
 \end{itemize}
\item Understanding the advanced stages of stellar evolution.
\item Probing the interstellar medium of our Galaxy.
\item Understanding the pulsar emission mechanism.
\end{itemize}

The aim of this paper is to investigate, for different configurations
of the SKA and different computation power and data rates, a variety
of different methods to achieve the best results for pulsar searches
and pulsar timing. We approach this problem by using Monte Carlo
simulations based on our current understanding of the Galactic pulsar
population. In \S 2, we summarise the various configurations for the
SKA. In \S 3, based on current population synthesis results, we
investigate the SKA survey yields as a function of configuration and
sky frequency. In \S 4, we describe a simple strategy for optimising
follow-up timing observations.  Computational requirements for the
pulsar surveys are discussed in \S 5, from which we make specific
recommendations in \S 6.  Finally, in \S 7, we summarise our results.

\section{SKA configurations}
We will express different sensitivities as a fraction of one `SKA
unit' defined as $2\times10^4$\,m$^2$\,K$^{-1}$. Following the
  preliminary specifications given by \citet{Schilizzi2007}, we
assume the SKA to consist of a sparse aperture array of tiled dipoles
in the frequency range of 70 to 500\,MHz and above 500\,MHz it
consists of one of the following three implementations:
\begin{itemize}
\item[A] 3\,000 15-m dishes with a single-pixel feed, a sensitivity of
  0.6 SKA units, $T_{\rm sys}$=30\,K and 70\% efficiency covering the
  frequency range of 500\,MHz to 10\,GHz.
\item[B] 2\,000 15-m dishes with phased array feeds from 500\,MHz to
  1.5\,GHz, a sensitivity of 0.35 SKA units, a field of view (FoV) of
  20\,deg$^2$, $T_{\rm sys}$=35\,K and 70\% efficiency and a
  single-pixel feed from 1.5 to 10\,GHz, with $T_{\rm sys}$=30\,K.
\item[C] A combination of a dense aperture array (AA) with a FoV of
  250\,deg$^2$, a sensitivity of 0.5 SKA units, covering the frequency
  range of 500 to 800\,MHz and 2\,400 15-m dishes with a single-pixel
  feed covering the frequency range of 800\,MHz to 10\,GHz, a
  sensitivity of 0.5 SKA units, $T_{\rm sys}$=30\,K and 70\%
  efficiency.
\end{itemize}

We assume that 20\% of the elements will be placed within a
1\,km radius and 50\% within a 5\,km radius. The sparse aperture
  array below 500\,MHz will be ignored, as we will not consider using
it for pulsar surveys or timing in this paper. We will refer to
  the dishes from configurations A and B (which have a lower frequency
  limit of 500\,MHz) as 500-MHz dishes and we will refer to the dishes
  from configuration C (which have a lower frequency limit of
  800\,MHz) as 800-MHz dishes.

\section{Pulsar surveys}

One of the main science goals for the SKA is to find the majority of
the pulsar population in the Galaxy. These should include binary and
millisecond pulsars, as well as transient and intermittent sources. A
pulsar survey consists of two parts. The first part is using the
telescope to observe parts of the sky with a certain dwell time. The
total amount of telescope time depends on the FoV of the telescope,
the chosen dwell time of the observation and the amount of sky that is
searched. The second part is the analysis of the observations, which
can be performed either real time, or off-line. Depending on the amount
of data that needs to be analysed and especially if acceleration
searches are employed to search for binary pulsars \citep[see
e.g.][]{lk05}, this can be a very computationally expensive task. Real
time processing has been done in the past, but any major pulsar survey
has always had an off line processing component \citep[see e.g.][]{Lorimer2006}.

\subsection{Technical considerations}

The FoV of a survey has a maximum that depends on the characteristics
of the elements used. In the case of circular dishes with a one pixel
receiver, the FoV can be approximated as $(\lambda/D_{\rm dish})^2$
steradians, where $\lambda$ is the wavelength at which the survey is
undertaken and $D_{\rm dish}$ is the dish diameter. However, by
placing a phased array feed at the focal point of the dish \citep[see
  e.g.][]{Ivashina04}, the FoV can be extended by a factor which we
denote $\xi$. The total FoV then no longer depends on $\lambda$. In
the case of the Aperture Array (AA), the FoV of the elements is about
half the sky. The actual size of the FoV that can be obtained will be
limited by the available computational resources. This is because the
signals from the elements will be combined coherently, resulting in
`pencil beams', the size of which scales with $1/D_{\rm el}^2$,
where $D_{\rm el}$ is the distance between the furthest
elements that are used in an observation. To obtain a large
synthesised FoV it is
therefore necessary to restrict the beam forming to using only the
elements in the core of the telescope (so that $D_{\rm el} =
  D_{\rm core}$), which will, however, reduce the sensitivity.

\subsection{Survey simulations}
To find out how the SKA pulsar survey performance depends on different
SKA configurations, we simulated SKA pulsar surveys for different
collecting areas and different centre frequencies. 

\subsubsection{Simulation method}
The simulations were performed following \citet{Lorimer2006} and using
their Monte Carlo simulation package\footnote{The package psrpop can
be obtained from \url{http://psrpop.sourceforge.net}.  Additional
software to simulate and analyse a series of surveys can be obtained
from: \url{http://www.jb.man.ac.uk/~rsmits/SKA.tgz}}.  In their
study, \citet{Lorimer2006} used the results from recent surveys with
the Parkes Multibeam system to derive an underlying population of
30,000 normal pulsars with an optimal set of probability density functions
(PDFs) for pulsar period ($P$), 1400-MHz radio luminosity ($L$),
Galactocentric radius ($R$) and height above the Galactic plane
($z$). We make use of these results in our simulations described below
which adopt the PDFs of model C' in \citet{Lorimer2006}.

Our simulation procedure begins, following the findings of
\citet{Lorimer2006}, by generating 30\,000 normal pulsars which beam
towards the Earth. Each pulsar is assigned a value of $P$, $L$, $R$
and $z$ based on the assumed PDFs. Their initial positions in the
plane of the Galaxy follow the spiral-arm modeling procedure described
by~\citet{Faucher2006}. The intrinsic pulse width of each pulsar
follows the power law relationship with spin period given in equation
(5) of~\citet{Lorimer2006} (see also their Figure 4). To compute the
expected DM and scatter broadening effects on each pulse, we use the
NE2001 electron density model~\citep{NE2001}. Note that, since we are
primarily concerned with the distant population of highly dispersed
pulsars in these simulations, we do not attempt to account for
interstellar scintillation. Finally, to allow us to extrapolate the
1400-MHz luminosities to other survey frequencies in the next step, we
make the reasonable assumption that pulsar spectra can be approximated
as a power law \citep{Lorimer95} and assign each pulsar a spectral
index drawn from a normal distribution with mean of --1.6 and standard
deviation 0.35. The same procedure was used to obtain a
  millisecond pulsar population, but with pulsar period and pulsar
  luminosity distributions from~\citet{Cordes97} and assuming 30\,000
  potentially detectable millisecond pulsars in the
  Galaxy~\citep{Lyne98}.

Given the above model for the underlying population, which is
consistent with the results of current-day surveys, we proceed to
apply it to different configurations of the SKA. For each
configuration, we compute the observed flux density of each pulsar and
compare it with the limiting flux threshold, which depends on the sky
temperature ($T_{\rm sky}$) and the observed width of the pulsar
profile. For the AA we reduced the sensitivity by the cosine of the
zenith angle of the source, as the AA is sensitive only to the
component of the radiation projected onto the plane of the AA.  In
addition, because the core of the SKA will be located near a
declination of --30$^\circ$, the sky visible to the SKA was limited to
a maximum declination of 50$^\circ$.

As discussed by \citet{Lorimer2006}, we make the important distinction
between the observed (i.e.~post-detection) pulse width which is
related to the pulsar's intrinsic pulse width by the quadrature sum of
contributions from sampling, dispersion smearing and scattering (see
equation 4 of \cite{Lorimer2006}). To scale the scatter-broadening
time to arbitrary SKA survey frequencies, we adopt a frequency power
law index of --4~\citep{NE2001}. We consider a pulsar to be detected
by a given configuration if its flux density exceeds the threshold
value, and its observed pulse width is less than the spin period.

\begin{figure}[htb]
\centering
\includegraphics[width=0.35\textwidth, angle=-90]{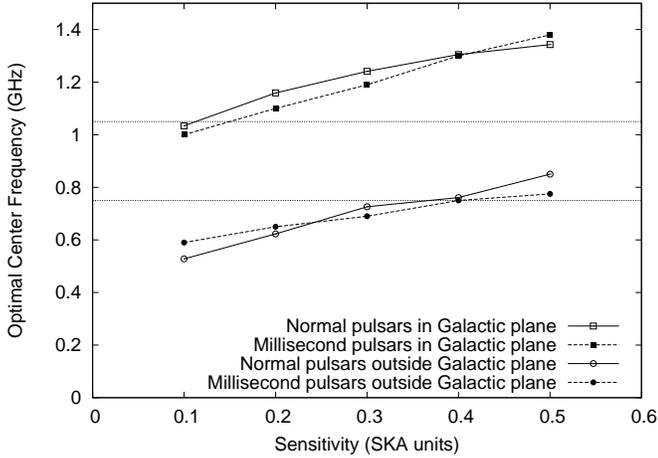}
\caption{Optimal frequency to survey the Galactic plane
  ($|$b$|<5^\circ$, $|$l$|<45^\circ$) or outside the Galactic plane
  ($|$b$|>5^\circ$, $|$l$|>45^\circ$) as a function of
  sensitivity. The dotted horizontal lines show the lowest available
  centre frequency of the 500-MHz (bottom line) and 800-MHz (top line)
  dishes. Surveys with the AA were performed at a centre frequency of
  650\,MHz.}
\label{fig:Optimal_frequency}
\end{figure}

Initially, we determined the optimal centre frequencies for a survey
in the Galactic plane (defined as $|$l$|<45^\circ$ and
$|$b$|<5^\circ$) and an all-sky survey excluding the Galactic
plane. This was done by comparing the detected number of normal
pulsars by simulating surveys with sensitivities in the range 0.1 to
0.5 SKA units (in steps of 0.1 SKA units) and with centre frequencies
in the range 0.4 to 2\,GHz (in steps of 0.25\,GHz). For each
sensitivity and for both survey types, the frequency that provided the
largest number of pulsars was chosen as the optimal frequency. To find
the frequency that provided the largest number of pulsars, a cubic
polynomial function was fitted through the points.  The optimal centre
frequencies for a pulsar survey inside and outside the Galactic plane
are shown in Fig.~\ref{fig:Optimal_frequency}. It can be seen that the
optimal survey frequency decreases with lower sensitivity. The reason
for this is that for lower sensitivity, the average distance to the
detected pulsar is smaller. Thus, frequency dependent propagation
effects are smaller, allowing lower frequencies to be used. It is
  surprising that the optimal centre frequencies for normal and
  millisecond pulsars are similar, since scattering effects on the
  pulsar profile have a larger impact on the detection of millisecond
  pulsars than on the detection of normal pulsars. Thus it can be
  expected that in the Galactic plane, higher observation frequencies
  would favour the detection of millisecond pulsars. However,
  millisecond pulsars have a steep luminosity
  distribution~\citep[e.g.][]{Cordes97}, leading to many
  low-luminosity millisecond pulsars which are not detected at high
  observation frequencies. It should be noted that with 500\,MHz
bandwidth, the lowest centre frequency is 750\,MHz for the 500-MHz
dishes and 1.05\,GHz for the 800-MHz dishes.  Once the optimal centre
frequencies were determined, these restrictions were included in the
simulations.
We then performed four simulations corresponding to possible pulsar
surveys. For the dishes they were performed at the optimal centre
frequencies, taking the frequency limits into account. The frequency
range of the AA was held fixed at 500 to 800\,MHz. The simulations are:
\begin{itemize}
\item An all-sky survey with the AA (including the Galactic plane) and
  a Galactic plane survey with the 800-MHz dishes. This shows the
  fraction of pulsars that can be found with configuration C.
\item An all-sky survey with just the AA. This shows the fraction of
  pulsars that will be found with configuration C, if we do not
  include the dishes in the survey.
\item An all-sky survey with the 500-MHz dishes. This shows the
  fraction of pulsars that can be found with configurations A or B.
\item A Galactic plane survey with the 500-MHz dishes.  This shows the
  fraction of pulsars that will be found with configuration A or B,
  without performing a survey outside the Galactic plane.
\end{itemize}
We assumed a bandwidth of 500\,MHz for the dishes and 300\,MHz for the
AA and an integration time of 1800\,s. The simulations were limited
to isolated pulsars only. A paper addressing the problems of the
searching for and timing of binary pulsars is in preparation (Smits et
al.~in prep.)

 A Galactic centre survey with the SKA will allow for testing the
  conditions near a $3\times10^6\,\mathrm{M}_\odot$ black
  hole. Simulating such a survey requires a detailed modelling of the
  scattering in the Galactic centre. Previous work
  \citep[e.g.][]{Lazio98, Bhat04, Cordes2004} shows that frequencies above 10\,GHz
  are required to penetrate the scattering screen surrounding the
  Galactic centre. These results can be used for a detailed study of a
  Galactic centre survey with the SKA. However, such a study is beyond
  the scope of this paper. Further results on this issue will be
  presented in J. Deneva et al. (in prep.)

\subsubsection{Simulation results}
\begin{figure}[htb]
\centering
\includegraphics[width=0.35\textwidth, angle=-90]{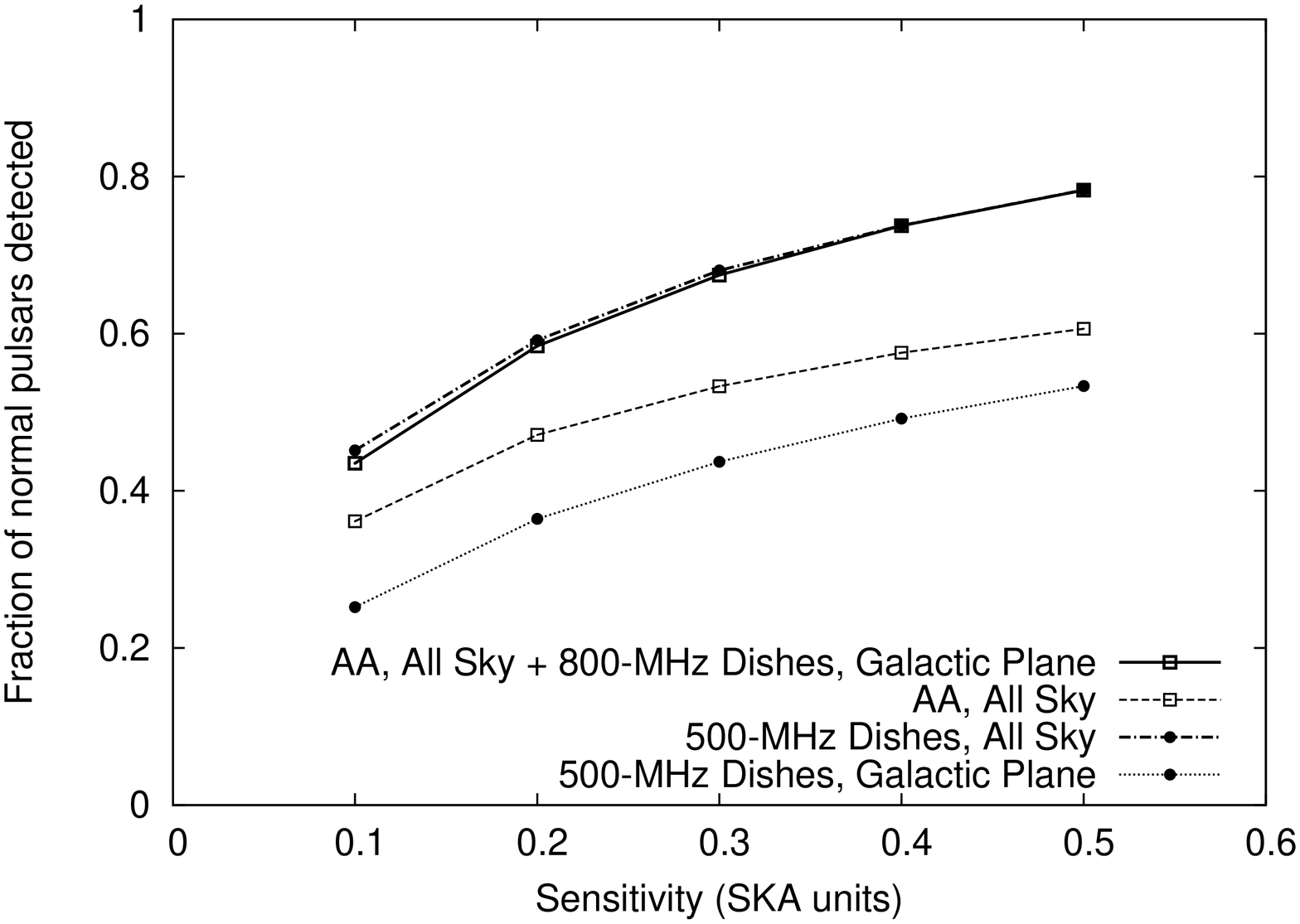}\\
\includegraphics[width=0.35\textwidth, angle=-90]{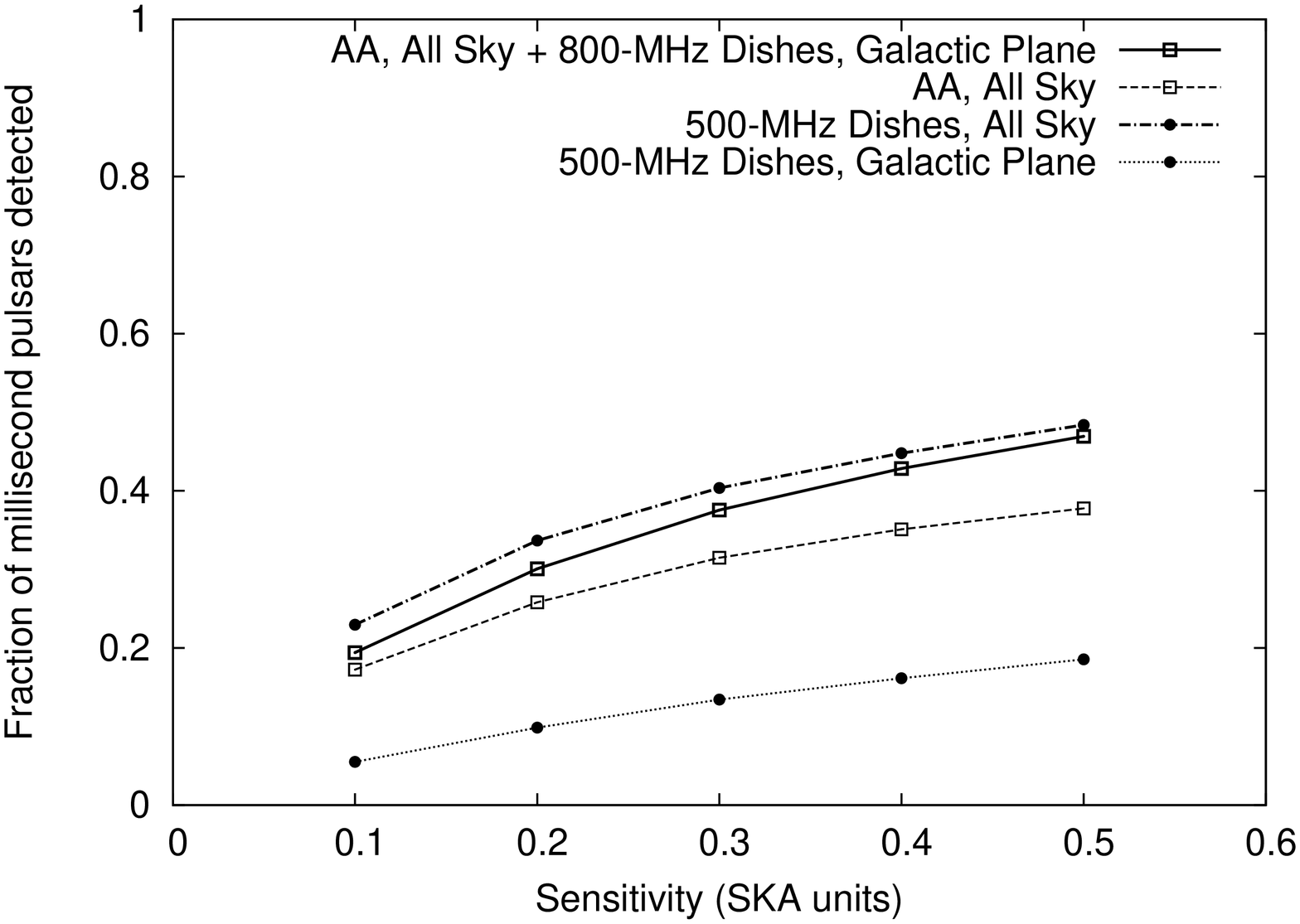}
\caption{Fraction of normal (top) and millisecond (bottom) pulsars
  detected from different pulsar survey simulations as a function of
  sensitivity of the SKA. For both the 500-MHz and 800-MHz dishes, the
  surveys were performed at the optimal centre frequency, taking the
  frequency limits into account (see
  Fig.~\ref{fig:Optimal_frequency}). The frequency range of the AA was
  500 to 800\,MHz.}
\label{fig:Simulation}
\end{figure}

 Fig.~\ref{fig:Simulation} shows the fraction of normal and
   millisecond pulsars that were detected in the simulation of 4
 different surveys as a function of sensitivity. Only 1.5\% of the
 pulsars could not be detected due to the declination of the SKA of
 --30$^\circ$. As will be shown in Section~\ref{sec:comp}, we expect
 that initially only the 1-km core of the SKA can be used for the
 pulsar survey. This leads to a sensitivity of 0.1 SKA units, which is
 similar to having a fully steerable Arecibo-class telescope in the
 southern hemisphere. Performing an all-sky survey with only the 1-km
 core of the SKA, will detect about 14\,000 normal pulsars out of a
 possible 30\,000 and about 6\,000 millisecond pulsars out of a
   possible 30\,000, regardless of the SKA implementation. The total
 observation time of the survey, however, does depend on the
 implementation. Using single-pixel feed dishes to perform an all-sky
 survey or a survey of just the Galactic plane will take 600 days and
 30 days, respectively. With phased array feed dishes or an AA the
 total survey time can be much less, depending on the available
 computation power (see Section~\ref{sec:suggestions}).  Because of
 the large natural FoV of the SKA and it's location in the southern
 hemisphere, the 1-km core of the SKA complements and extends work
 done by both Arecibo and the Five hundred meter Aperture Spherical
 Telescope (FAST), which is expected to be completed around 2014.
\section{Timing}
\begin{figure}[htb]
\centering
\includegraphics[width=0.35\textwidth, angle=-90]{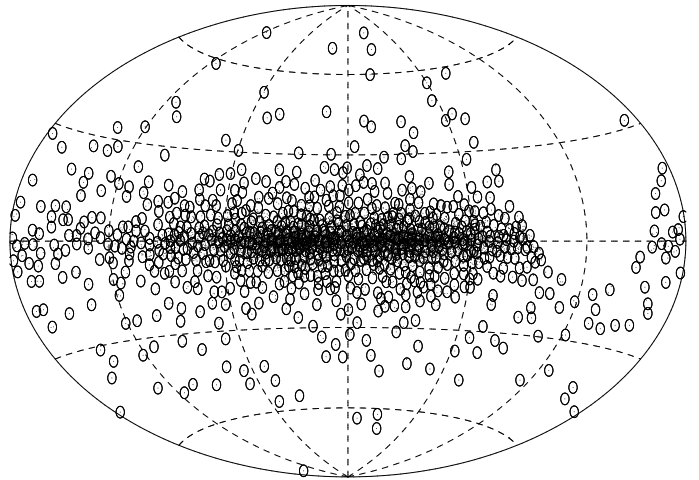}
\caption{Hammer projection of the Galaxy with 14\,000 detected pulsars
 from an SKA pulsar survey simulation. The horizontal axis corresponds
 to longitude and the vertical axis corresponds to latitude. The
 Galactic centre is located in the middle of the plot. The dashed
 lines running from top to bottom correspond to steps of 60$^\circ$ in
 longitude, whereas the dashed lines running from left to right correspond to
 steps of 30$^\circ$ in latitude. The circles indicate all
 pointings (700) that result from the timing optimisation method,
 assuming a FoV of 20\,deg$^2$.}
\label{fig:Hammer}
\end{figure}

All pulsars that will be found by the SKA need to become part of a
regular timing program to find the interesting pulsars. Ideally, they
need to be timed once every two weeks for about 6 months to a
signal-to-noise ratio of about 9 (which is only a nominal target to
characterise the new pulsars and to identify those sources which are
interesting for high precision timing)\footnote{Usually, at least 12
  months are needed to break the degeneracies between positional
  uncertainties and pulsar spin-down.  However, we assume that the
  imaging capabilities of the SKA will be used after the discovery to
  obtain an interferometric position, which speeds up the process.}.
 The minimum timing duration is also restricted by the pulsar
  stabilisation time, which is the time it takes to obtain a stable
  integrated profile.  Assuming that they will be timed using dishes,
the total duration of timing all newly discovered pulsars just once
depends on how the observations are performed, i.e.~at which points on
the sky the telescope is pointed. To estimate the total duration of
this timing process, we therefore suggest the following scheme that
minimises the average observation time per pointing, by grouping dim
pulsars together in one FoV.  Let $\rho_{\rm fov}$ be the angle of the
FoV.
\begin{enumerate}
\item All pulsars are placed on a long list, in order of increasing brightness.
\item All pulsars within an angle of $\rho_{\rm fov}$ from the first pulsar of the list (the dimmest pulsar) are placed on a short list.
\item We now consider the dimmest pulsar on the short list. If
 possible, we restrict the pointing of the FOV to include this pulsar and
 we remove this pulsar from both the short list and the long list. If
 the pointing of the FOV is already too restricted to include this
 pulsar, it is removed from the short list only.
\item Step 3 is repeated until the short list is empty. The resulting pointing is stored.
\item Step 2 through 4 are repeated until the long list is empty.
\end{enumerate}
This timing optimisation method saves about 30 to 50\% of observation
time over a simple grid approach. To compare timing performances, we
used this scheme to estimate the time to obtain a single timing
  point of 14\,000 pulsars from an all-sky survey to a
signal-to-noise ratio of 9 for each of the SKA-configurations,
assuming usage of the full collecting area and a pulsar stabilisation
time~\citep[see e.g.][]{Helfand75} of 200 pulses. This leads to
observation times of 6 days for configuration A and 1.5 days for
  both configurations B and C. For configuration C the AA was used to
time all the pulsars that were detected by the AA in the survey. These
estimates do not include densely spaced `resolving' observations that
are required to obtain the first phase coherent timing solution.
Fig.~\ref{fig:Hammer} shows the simulated pulsar-sky and the pointings
that result from the timing optimisation method, assuming the
20\,deg$^2$ FoV of configuration B.

 To perform strong field tests of General Relativity and gravitational
 wave detection, as defined in the key science project, it is
 essential to time a large number of specific pulsars to
 signal-to-noise ratios of up to 100 on a regular basis. To estimate
 how much observation time is required for such high-precision timing,
 we have to distinguish between timing millisecond pulsars (for the
 Pulsar Timing Array) and timing pulsars in a binary with a neutron
 star or a black hole companion. We calculated the observation time
 for timing the 250 millisecond pulsars with the best signal-to-noise
 ratio's for the three configurations. We assumed that for timing
 purposes, the nearly full collecting area (out to several hundreds of
 kilometres from the core) of the SKA can be used. As a conservative
 estimate (subject to current studies), we timed every pulsar for at
 least 5 minutes to ensure a stable pulsar profile to minimise the
 error in the time of arrival. With the single-pixel feed 15-m dishes
 this takes about 20 hours, the 15-m dishes with phased array feeds
 take about 15 hours and the AA only takes about 6 hours. However,
 this assumes that the polarisation purity after calibration of the AA
 is similar to that of the dishes, which might not be the case. To
 estimate the maximum required observation time for timing pulsars in
 a binary with a neutron star or black hole, we will assume a
 (somewhat optimistic) number of 200 such binaries that are
 potentially detectable in the Galaxy. We further assume that the
 characteristics of the pulsars in these binaries are similar to
 isolated pulsars. This leads to a detection of 90 compact binary
 pulsars. With the single-pixel feed 15-m dishes it takes about 2 days
 to time these binary pulsars, the 15-m dishes with phased array feeds
 take about 1.5 days and the AA only takes about 8 hours (once again,
 assuming similar polarisation purity as the dishes). However, timing
 only the brightest 80\% of the binary pulsars, would take about half
 the observation time. It should be noted that these are only initial
 estimates to give an indication of the maximum required observation
 time for high-precision timing and to compare the performance of the
 different configurations. A paper containing a more detailed study on
 high-precision timing is in preparation (Smits et al.~in prep.).

\section{Computational requirements}
\label{sec:comp}
A pulsar survey requires the coherent addition of the signals from the
elements in the core of the SKA and to form sufficient pencil beams to
create the required FoV. These pencil beams will produce a large
amount of data which will need to be analysed. We estimate the
required computation power and data rates for 3\,000 15-m dishes with
a bandwidth of 500\,MHz and 2 polarisations and for the AA with a
frequency range from 500 to 800\,MHz and 2 polarisations.

\subsection{Beam forming}
\label{sec:Beamforming}
\begin{figure}
 \centering
 \includegraphics[width=0.35\textwidth, angle=-90]{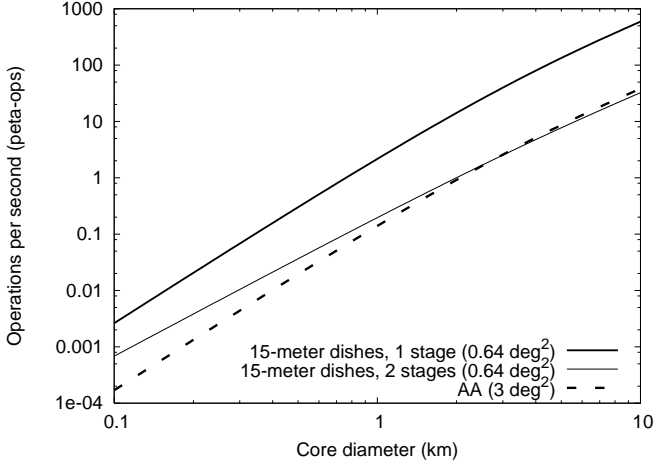} 
\caption{The number of operations per second required to perform the
  beam forming for the 15-m dishes and the AA. It is assumed that
  there are 2\,400 15-m dishes and the bandwidth is 500\,MHz. The FoV
  of 0.64\,deg$^2$ corresponds to filling the natural FoV of the
  dishes ($\xi=1$) at a frequency of 1.4\,GHz. For the AA we assume a
  FoV of 3\,deg$^2$ and a total collecting area of 500\,000\,m$^2$.
  The frequency range is 0.5 to 0.8\,GHz. In all cases the number of
  polarisations is 2. The thick black line corresponds to the beam
  forming of the 15-m dishes in one stage. The thin black line
  corresponds to the beam forming of the 15-m dishes in two
  stages. The striped/dotted line corresponds to the beam forming of
  the AA.}
 \label{fig:Beamforming}
\end{figure}

Following~\citet{Cordes07}, we estimate the number of operations per
second (ops), $N_{\rm osb}$, to fill the entire FoV of a dish with pencil
beams as\footnote{Note that for a phased array feed this function is
actually frequency dependent, because the total FoV of the dish then
becomes constant, yet the FoV of the pencil beams is given by
$(\lambda/D_{\rm tel})^2$. However, for the values of bandwidth and
observation frequency used in this paper, Eq.~\ref{eq:beamforming} is
accurate.}:
\begin{equation}
 N_{\rm osb} = F_c N_{\rm dish} N_{\rm pol} B \xi\left(\frac{D_{\rm core}}{D_{\rm dish}}\right)^2,
 \label{eq:beamforming}
\end{equation}
where $F_c$ is the fraction of dishes inside the core, $N_{\rm dish}$
is the total number of dishes, $N_{\rm pol}$ is the number of
polarisations, $B$ is the bandwidth, $D_{\rm dish}$ is the diameter of
the dishes, $D_{\rm core}$ is the diameter of the core and $\xi$ is
the factor by which the phased array feed extends the FoV.  To fill
the FoV of single-pixel 15-m dishes in a 1-km core requires 4\,400
pencil beam. The number of operations per second to beam form a 1-km
core containing 20\% of 2\,400 single-pixel 15-m dishes then becomes
$2.2\times10^{15}$. The number of operations for beam forming the full
FoV of 15-m dishes with a phased array feed are simply a factor $\xi$
higher, corresponding to the increase in FoV. For the SKA
implementation B, this corresponds to $\xi=31$. The number of
operations for beam forming can be reduced by beam forming in two
stages, if we assume that the dishes in the core are positioned such
that they can form sub-arrays that all have the same size. In the
first stage, beam forming of the full FoV is performed for each
sub-array. In the second stage the final pencil beams are formed by
coherently adding the corresponding beams of each sub-array formed in
the first stage. When the sub-arrays have the optimal size given by
$D_{\rm core}(F_c N_{\rm dish})^{-1/4}$, as we show in the Appendix,
the number of operations per second becomes
\begin{equation}
N_{\rm osb} = 2 \sqrt{F_c N_{\rm dish}} N_{\rm pol} B \xi \left(\frac{D_{\rm core}}{D_{\rm dish}}\right)^2.
\end{equation}
For the same parameters as above, the number of operations per second becomes
$2.0\times10^{14}$. For the beam forming of the AA, the calculations are slightly
different. Initial beam forming will be performed at the stations themselves,
leading to beams equivalent to those of a 60-m dish for each
station. Demanding a total FoV of 3\,deg$^2$ and assuming a total collecting
area of 500\,000\,m$^2$, the number of operations becomes:
\begin{equation}
 N_{\rm osb} = F_c N_{\rm dish} N_{\rm pol} B\cdot3\left(\frac{\pi}{180c}\right)^2 D_{\rm core}^2\nu_{\rm max}^2,
\label{eq:beamformingAA}
\end{equation}
where $c$ is the speed of light and $N_{\rm dish} \approx 177$.  For a
1-km core, this leads to a required computation power of
$1.4\times10^{14}$\,ops.  Fig.~\ref{fig:Beamforming} shows the number
of operations per second for the beam forming as a function of core
diameter for the 15-m dishes with single-pixel feed and the AA. To
obtain the benchmark concentration as a continuous function of core
diameter between 0 and 10~km, we used the following expression:
\begin{equation}
F_c(D_{\rm core}) = a[1-\exp(-bD_{\rm core})],
\label{eq:F_c}
\end{equation}
where $a$ and $b$ were tuned to the specifications of 20\% and 50\% of the
dishes within a 1-km and 5-km core, respectively, which leads to $a=0.56$ and
$b=0.45\times10^{-3}$\,m$^{-1}$.

As an alternative to beam forming by coherently adding the signals
from the dishes up to a certain core size, it is also possible to
perform the beam forming by incoherently adding the signals from
sub-arrays. This process is similar to beam forming in two stages as
mentioned above, except that in the second stage the beams that were
formed in the first stage are added incoherently. This leads to a much
larger beam size, which reduces the required computation power for
beam forming significantly. It also reduces the total data rate and
the required computation power for the data analysis. The drawback is
that it reduces the sensitivity of the telescope by a factor of
$\sqrt{N_{\rm sa}}$, where $N_{\rm sa}$ is the number of sub-arrays,
which will be several hundred. However, this can be partially
compensated as this method allows utilising all elements that are
placed in sub-arrays of the same size, which can possibly be as much
as 5 times the collecting area of the core of the SKA.

\subsubsection{Data analysis}
\label{sec:data analysis}
There are two factors which impact the total data volume to be
analysed.  Firstly, the FoV will be split up in many pencil beams,
each of which will need to be searched for pulsar signals. Secondly,
the SKA will be able to see the majority of the sky, all of which we
want to search for pulsars and pulsar binaries. There are two ways to
achieve this. The first option is to analyse the data as it is
received, immediately discarding the raw data after analysis. This
requires the analysis to take place in real time. The second option is
to store all the data from part of the survey and analyse them at any
pace that we see fit before continuing with the next part of the
survey. Both approaches pose serious technical challenges, which we
will discuss here.

\begin{figure}
 \centering
 \includegraphics[width=0.35\textwidth, angle=-90]{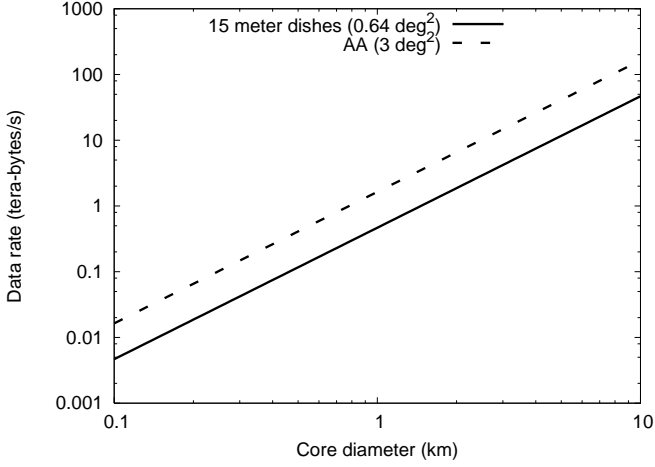}
 \caption{Data rate from pulsar surveys using the 15-m dishes or the AA, as a function of core
 diameter.}
 \label{fig:Datarate}
\end{figure}

\begin{figure}
 \centering
 \includegraphics[width=0.35\textwidth, angle=-90]{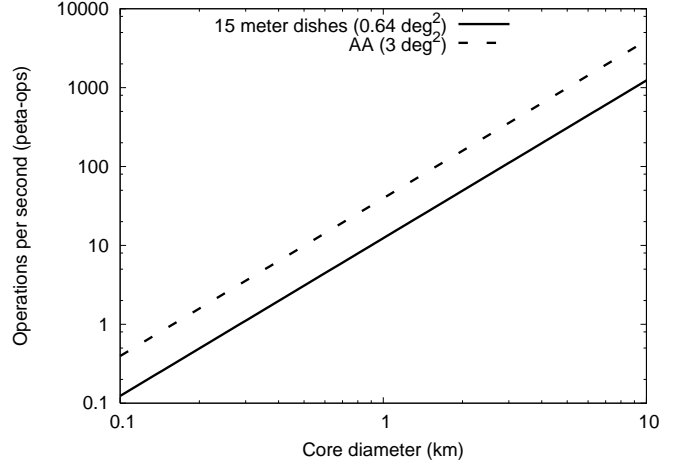}
 \caption{Operations per second required to perform a real time analysis of pulsar surveys using the 15-m dishes or
 the AA, as a function of core diameter.}
 \label{fig:Analysis}
\end{figure}

First we consider the dishes for which we estimate the data rate of one pencil beam as
\begin{equation}
{\cal D}_{\rm dish} = 
 \frac{1}{t_{\rm samp}}\frac{B}{\Delta\nu}N_{\rm pol}\frac{N_{\rm bits}}{8}\mathrm{\,Bps},
 \label{eq:DR}
\end{equation}
where $t_{\rm samp}$ is the sampling time, $B$ is the bandwidth, $\Delta\nu$ is
the frequency channel width, $N_{\rm pol}$ is the number of polarisations and
$N_{\rm bits}$ is the number of bits used in the digitisation. $\Delta\nu$ can be
estimated by demanding that the dispersion smearing within the frequency
channel does not exceed the sampling time, which leads to:
\begin{equation}
 \Delta\nu(\mathrm{GHz}) = \frac{t_{\rm samp}(\mu\mathrm{s})\nu_{\rm min}^3(\mathrm{GHz})}{8.3\times10^3{\rm DM}_{\rm max}},
\end{equation}
where $\nu_{\rm min}$ is the minimum (lowest) frequency in the observation frequency band and DM$_{\rm max}$ is the maximum
expected dispersion measure. For the dishes, the number of pencil beams to fill up the FoV can be estimated as
$\xi(D_{\rm core}/D_{\rm dish})^2$. For the AA, the number of pencil beams becomes frequency dependent and for 3\,deg$^2$
FoV it is given by
\begin{equation}
N_{\rm pbAA} = 3\left(\frac{\pi}{180c}\right)^2 D_{\rm core}^2\nu^2.
\label{eq:npbAA}
\end{equation}
However, a pencil beam needs to be centred
on the same point on the sky over the frequency band to allow for summing over
frequency. Thus, even though the pencil beams become larger towards lower
frequency, the number of pencil beams needs to remain constant and is
determined by the highest frequency. The total data rate from the AA can then
be estimated as
\begin{equation}
{\cal D}_{\rm AA} = 
\frac{1}{t_{\rm samp}}\frac{B}{\Delta\nu}N_{\rm pol}\frac{N_{\rm bits}}{8} 3\left(\frac{\pi}{180c}\right)^2
 D_{\rm core}^2\nu_{\rm max}^2 \mathrm{\,Bps}.
\end{equation}
Fig.~\ref{fig:Datarate} shows the data rate from a pulsar survey for the
15-m dishes and the AA as a function of core diameter, assuming
$t_{\rm samp}$=100\,$\mu$s, DM$_{\rm max}$=1000\,cm$^{-3}$\,pc 
for the dishes and DM$_{\rm max}$=500\,cm$^{-3}$\,pc for
the AA, $N_{\rm pol}$=1 (sum of 2 polarisations) and 2 bit digitisation. 
The AA was assumed to operate on
the full frequency range of 0.5 to 0.8\,GHz. The frequency range for the
dishes was assumed to be 1 to 1.5\,GHz. For a 1-km core, the data rate for the
15-m dishes with single-pixel feed is $4.7\times10^{11}$\,bytes per second
and for the AA it is $1.6\times10^{12}$\,bytes per second. Both values scale
linearly with the FoV.
We can estimate the number of operations to search one `pencil beam' for
accelerated periodic sources as one Fourier Transform of all the samples in
the observation, for each trial DM-value and for each trial acceleration:
\begin{equation}
 N_{\rm oa}=N_{\rm DM}N_{\rm acc}\times5N_{\rm samp}\log_2(N_{\rm samp}),
\end{equation}
where $N_{\rm DM}$ is the number of trial DM-values, $N_{\rm acc}$ is the
number of trial accelerations which scales with the square of
$N_{\rm samp}$, which is the number of samples in one observation. Thus,
the number of operations for the analysis scales as
$N_{\rm samp}^3\log(N_{\rm samp})$. This means that increasing the observation
length is computationally very expensive. Once again, for the dishes
the number of pencil beams to fill up the FoV can be estimated as
$\xi(D_{\rm core}/D_{\rm dish})^2$ and for the AA the number of pencil
beams is given by Eq.~\ref{eq:npbAA}, with
$\nu$=$\nu_{\rm max}$=0.8\,GHz. Fig.~\ref{fig:Analysis} shows the number
of operations per second required to perform a real time analysis of a
pulsar survey with the 15-m dishes and the AA as a function of
core diameter, assuming 100 trial accelerations, a sampling time of
100\,$\mu$s and an observation time of 1\,800\,s. $N_{\rm DM}$ was taken
to be equal to the time shift due to the maximum dispersion divided by
the sampling time, which is equal to
\begin{equation}
N_{\rm DM} = \frac{4150 DM_{\rm max}(\nu^{-2}_{\rm min}({\rm
  GHz})-\nu^{-2}_{\rm max}({\rm GHz}))}{t_{\rm samp}({\rm \mu s})}. 
\end{equation}
For a 1-km core, the required computation power becomes
1.2$\times10^{16}$ and 4.0$\times10^{16}$\,ops for the 15-m dishes and
the AA, respectively. These values are lower estimates of the
  actual number of operations, as there will be significant
  contributions from dedispersion, harmonic folding and possibly other
  processes that will contribute to the total number of operations.
  These additional processes are all on the order of $N_{\rm samp}$
  and we expect their total contribution to be similar to one Fourier
  Transform at most.

\begin{figure}
 \centering
 \includegraphics[width=0.35\textwidth, angle=-90]{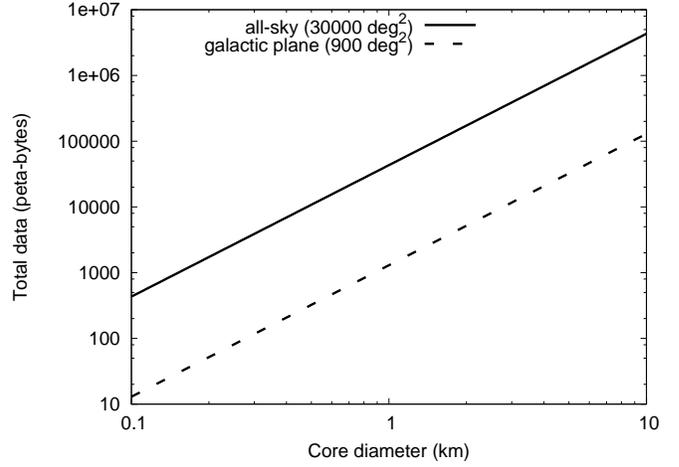}
 \caption{Total amount of data from an all-sky survey (35\,000\,deg$^2$) and a survey of the Galactic plane
 (900\,deg$^2$) as a function of core diameter, assuming a frequency
 range of 1 to 1.5\,GHz. A 1-km core would correspond to 20\% of the
 collecting area of the SKA.}
 \label{fig:Data}
\end{figure}

Alternatively, we can consider storing the data from a part of the
survey and analyse it at a much slower rate (see further discussion in
Section~\ref{sec:offline analysis}). From Eq.~\ref{eq:DR} we can
estimate the total amount of data from a survey. Fig.~\ref{fig:Data}
shows the total amount of data from an all-sky survey and a survey of
the Galactic plane as a function of core diameter, assuming a
frequency range of 1 to 1.5\,GHz, $t_{\rm samp}$=100\,$\mu$s, DM$_{\rm
max}$=1000\,cm$^{-3}$\,pc, $N_{\rm pol}$=1, an observation time of
1\,800\,s and 2 bits digitisation. Assuming a storage capacity of 1
exa-byte, an all sky survey would have to be split up in about 40
parts. The total time of the survey would then depend on the speed at
which the analysis takes place.
\subsubsection{Survey speed}
\label{sec:survey_speed}
To express the trade-off between the SKA design and survey speed, we define the figure of merit for dishes as (see
appendix for derivation)
\begin{equation}
 M = \eta \xi B\left(\frac{N_{\rm dish} D_{\rm dish}}{T_{\rm tel}+T_{\rm sky}^{400}\left(\frac{\nu}{400\mathrm{\,MHz}}\right)^{-2.6}}\right)^2\nu^{2\alpha-2},
\end{equation}
where $\eta$ is the aperture efficiency, $\xi$ is the factor by which
the phased array feed extends the FoV, $B$ is the bandwidth, $N_{\rm
  dish}$ is the number of dishes used in the survey, $D_{\rm
  dish}$ is the diameter of the dish, $T_{\rm tel}$ is the system
temperature of the telescope (K), $T_{\rm sky}^{400}$ is the sky
temperature (K) at 400\,MHz ($T_{\rm sky}$ scales as $\nu^{-2.6}$),
$\nu$ is the observation frequency, and $\alpha$ is the average
spectral index of radio pulsars, which is about --1.6
\citep{Lorimer95}. The figure of merit shown here expresses the
  speed of the survey for a required sensitivity of the telescope and
  is identical to that for a continuum survey of steady sources. It
  does not include the loss of sensitivity associated with pulse
  broadening effects and does not express the impact of dedispersion
  and periodicity search on the survey time.

\begin{figure}
\centering
\includegraphics[width=0.35\textwidth, angle=-90]{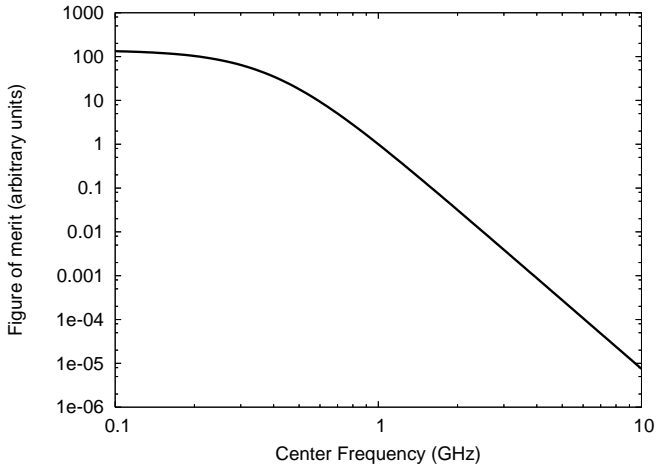}
\caption{The frequency dependence of the figure of merit. This function expresses the speed of the survey for a fixed
 sensitivity of the telescope.}
\label{fig:Merit}
\end{figure}
Fig.~\ref{fig:Merit} shows the strong dependence of the figure of merit on the
observation frequency. The slope at high frequencies is due to the spectral
index of pulsars and the decrease of the FoV. At low frequency the figure of
merit becomes flat as the sky temperature begins to dominate the system
temperature. In practise, pulsars become harder to detect at low frequencies
where scattering effects broaden the pulsar profile. Since the scattering
depends on the location on the sky, it was not included in the merit
function. Thus, at low frequencies the figure of merit is
overestimated. However, scattering effects are only significant for pulsars
located in the Galactic plane. For non-millisecond pulsars, low frequencies are
very favourable~\citep[see][]{Leeuwen08}.

The total observation time of a survey is given by:
\begin{equation}
 t_{\rm total} = \frac{\Omega_{\rm sur}}{\Omega_{\rm FoV}}t_{\rm point}
 \label{eq:survey_speed}
\end{equation}
where $\Omega_{\rm sur}$ is the total solid angle of sky covered
by the survey, $\Omega_{\rm FoV}$ is the total FoV of the dishes and $t_{\rm point}$ is
the observation time per pointing.

\section{Suggestions}
\label{sec:suggestions}
An estimate of the computational power by 2015 is given by~\cite{Cornwell2005}
to be 10\,Pflop for \$100\,M. To estimate what the SKA can achieve for pulsar
searches and timing, we will assume that several $10^{16}$\,ops are available
for beam forming and data analysis when this survey will be performed, which
will be after 2020.

We consider three scenarios, two with full coherent beam forming, where the
analysis takes place either in real time or off-line and one scenario where part
of the beam forming takes place incoherently.

\subsection{Real time analysis}

First we assume that the analysis takes place real time. The
immediate benefits are that no time is lost for the data analysis and
that only reduced data needs to be stored. One drawback, however, is
that the data can only be processed once. Experience with current
pulsar surveys show that it is extremely advantageous to have multiple
passes at pulsar survey data analysis \citep[see e.g.][]{Faulkner05}.

As calculated in Section~\ref{sec:data analysis}, for a 1-km core the
data rate from the dishes with a single-pixel feed is
$4.7\times10^{11}$\,bytes/s. The required computational power for real
time analysis is $1.2\times10^{16}$\,ops. Both values scale linearly
with FoV. Only with SKA implementation B, where the dishes have phased
array feeds, can the FoV of the dishes be increased beyond
0.64\,deg$^2$. Using the 1-km core of the AA and using a FoV of
3\,deg$^2$ leads to a data rate of $1.6\times10^{12}$\,bytes/s and
requires a computation power of $4.0\times10^{16}$\,ops to analyse the
data in real time.

SKA implementation A suggests the all-sky survey to be performed with
single-pixel 15-m dishes. At low frequencies the FoV of the dishes is
about 1\,deg$^2$. The entire survey will then take about 600
days. (This value assumes 100\% telescope time. In practice this
  survey might take up to 5 years to perform). In implementation B a
larger FoV can be used. The survey time of 600 days then scales
inversely with the FoV, however the data rates and required
computation power go up linearly with the FoV. Implementation C has
the benefit of the AA for which an all-sky survey takes only about 200
days and the survey of the Galactic plane with single-pixel 15-m
dishes takes about 30 days. The data rates and required computation
power for the AA are larger than for the dishes. When we take this
into account the total survey time for all implementations become
similar, except that implementations B and C allow for a faster survey
and the AA in implementation C might be used for other observations
simultaneously~\citep[see][for using the SKA as a synoptic survey
  telescope]{Cordes07}.

\subsection{Off-line analysis}
\label{sec:offline analysis}
The off-line analysis requires the full data from an observation to be
stored. Thus, the data rates from Fig.~\ref{fig:Datarate} become the rates at
which data is written to a storage device. When the maximum data storage has
been reached, the observation can be stopped and the data analysis can be
performed at any pace suitable. After the data analysis has been completed,
the next observation can be run. This allows a trade-off between computation
cost and survey speed. A drawback is that, depending on available future
  storage solutions, the total survey time may easily become more than a
decade for any SKA implementation.

\subsection{Incoherent beam forming}
For real time analysis, restrictions on the computational power and data rate
will likely limit the FoV to about 3\,deg$^2$. The survey time will then
become at least 200 days. Off-line analysis requires the huge data rates to be
written to a storage device and will increase the survey time by a significant
factor. We therefore consider the possibility of incoherent addition of the
sub arrays, as described in Section~\ref{sec:Beamforming}.

We will assume that all the elements in the SKA are placed such that
their signals can be combined to form sub-arrays (or stations in the
case of the AA) of 60 metres in diameter~\citep[similar to][for
  LOFAR]{Leeuwen08}. The pencil beams that are formed by coherently
adding the signals from the sub-array elements are about 17 times
larger than the pencil beams from coherently beam forming the 1-km
core. Thus the number of pencil beams required to fill the FoV becomes
about 280 times smaller. The computational requirements for beam
forming, data analysis as well as the data rates go down linearly by
this factor. This method of beam forming utilises the full collecting
area of the SKA, which increases the sensitivity by a factor of
5. However, because the sub-arrays are added incoherently, the
sensitivity decreases by the square root of the number of
sub-arrays. For the AA there will be about 180 stations. Thus the
sensitivity goes down by a factor of
$\sqrt{180}/5\approx2.7$. Assuming that the dishes are placed in
sub-arrays with a 20\% filling factor, the number of sub-arrays will
be between 600 and 900, depending on the SKA implementation. For
incoherent beam forming the sensitivity will then go down by a factor
between 5 and 6. Applying this to an all-sky survey leads to a
  detection of about 20\% of all the detectable pulsars.

\section{Summary}

We have investigated the pulsar yields for different SKA
configurations by simulating SKA pulsar surveys for different
collecting areas and centre frequencies. For the Galactic plane, the
optimal centre frequency lies just above 1\,GHz. Outside the Galactic
plane the optimal centre frequency lies between 600 and 900\,MHz,
depending on the collecting area. Combining the dishes and the AA to
perform a survey inside the Galactic plane and outside the Galactic
plane, respectively, would result in detection of roughly 14\,000
  normal pulsars and 6\,000 millisecond pulsars. As
Fig.~\ref{fig:Simulation} shows, increasing the collecting area within
the 1-km core would improve the detected number of pulsars
significantly. The number of detected pulsars scales roughly as
$F_c^{0.4}$, where $F_c$ is the fraction of collecting area in the
core.

We also describe a simple strategy for follow-up timing observations
and find that, depending on the configuration, it would take 1--6 days
to obtain a single timing point for 14\,000 normal
pulsars. Configuration C, containing the AA, will be of great benefit
here. Obtaining a single timing point for the high-precision timing
projects of the SKA, will take less than 14 hours, 2 days, or 3 days,
depending on the configuration. Once again, the presence of the AA
allows for the fastest timing. In the case of high-precision timing,
however, this assumes that the polarisation purity of the AA is
similar to that of the dishes.

Performing a pulsar survey with the SKA requires the coherent addition
of the signals of the individual elements, forming sufficient pencil
beams as to fill the entire FoV of a single dish. Because of the
extreme computational requirements that arise due to large
baselines, it is not possible to combine the signals of all of the
elements in the SKA. Rather, only the core of the SKA can be used for
a pulsar survey. We have derived the computational requirements to
perform such beam forming as a function of core-diameter for the 15-m
dishes and the AA. For a 1-km core the requirements are
2.2$\times10^{15}$ and 1.4$\times10^{14}$\,ops for the 15-m dishes and
the AA, respectively. When the dishes are placed such that they can form
sub-arrays, the computational requirement for beam forming
goes down significantly. We have also calculated the data rates and the
computational requirements for applying a search-algorithm to the data
to find binary pulsars. Both limit the usage of the SKA for pulsar
searches to a core of about 1~km.

\acknowledgement{ The authors would like to thank Simon Johnston and
  Andrew Lyne for their useful suggestions and the referee, Scott
  Ransom, for his useful suggestions and comments. This
  effort/activity is supported by the European Community Framework
  Programme 6, Square Kilometre Array Design Studies (SKADS), contract
  no 011938. DRL is supported by a Research Challenge Grant from West
  Virginia EPSCoR. Research at Cornell University was supported by NSF
  Grant AST0507747.}

\section*{Appendix}
\subsection*{Beam forming of the dishes}

Cordes (2007) estimated the number of operations to perform the beam forming
for a pulsar survey as one operation (corresponding to the phase shift of one
element) for each dish and each polarisation at the Nyquist frequency. This
needs to be performed for each of the pencil beams that fill the total
FoV. This leads to the total number of operations per second
\begin{equation}
 N_{\rm osb} = F_c N_{\rm dish} N_{\rm pol} B \left(\frac{D_{\rm core}}{D_{\rm dish}}\right)^2,
 \label{eq:beamforming2}
\end{equation}
where $F_c$ is the fraction of dishes inside the core, $N_{\rm dish}$ is the total
number of dishes, $N_{\rm pol}$ is the number of polarisations, $B$ is the
bandwidth, $D_{\rm dish}$ is the diameter of the dishes and $D_{\rm core}$ is the
diameter of the core. The fraction $(D_{\rm core}/D_{\rm dish})^2$ is equal to the
required number of pencil beams to fill up the FoV. If we assume that the
dishes in the core are positioned such that they can form sub-arrays,
beam forming can take place in two stages. In the first stage beam forming of the
full FoV is performed for each sub-array. In the second stage the final pencil
beams are formed by coherently adding the corresponding beams of each
sub-array formed in the first stage. This leads to the following expression
for the number of operations required for beam forming in two stages:
\begin{eqnarray}
 N_{\rm osb2} & = & N_{\rm sa}N_{\rm dishsa} N_{\rm pol} B\left(\frac{D_{\rm sa}}{D_{\rm dish}}\right)^2 \nonumber \\
 & & + \left(\frac{D_{\rm sa}}{D_{\rm dish}}\right)^2 
N_{\rm sa} N_{\rm pol} B\left(\frac{D_{\rm core}}{D_{\rm sa}}\right)^2,
\end{eqnarray}
where $N_{\rm sa}$ is the number of sub-arrays in the core, $N_{\rm dishsa}$ is the
number of dishes in one sub-array and $D_{\rm sa}$ is the diameter of the
sub-arrays. Substituting $N_{\rm dishsa} = F_c N_{\rm dish}/N_{\rm sa}$ and $N_{\rm sa} =
(D_{\rm core}/D_{\rm sa})^2$ yields
\begin{eqnarray}
 N_{\rm osb2} & = & F_c N_{\rm dish} N_{\rm pol} B \left(\frac{D_{\rm sa}}{D_{\rm dish}}\right)^2 \nonumber \\
 & & + N_{\rm pol} B \left(\frac{D_{\rm core}}{D_{\rm dish}}\right)^2 
\left(\frac{D_{\rm core}}{D_{\rm sa}}\right)^2,
\end{eqnarray}
which has a minimum at $D_{\rm sa} = D_{\rm core}(F_c N_{\rm dish})^{-1/4}$. At this
minimum, the number of operations for beam forming in 2 stages is:
\begin{equation}
N^{\rm min}_{\rm osb2} = 2 \sqrt{F_c N_{\rm dish}} N_{\rm pol} B \left(\frac{D_{\rm core}}{D_{\rm dish}}\right)^2.
\end{equation}

\subsection*{Beam forming of the AA}
For the AA, initial beam forming will be performed at each station; this will
lead to beams equivalent to those of a 60-m dish.  Assuming a collecting area
of 500\,000\,m$^2$, this leads to about 177 stations. The FoV of 1 pencil beam
is given by $(\lambda/D_{\rm Core})^2 = (c/\nu D_{\rm core})^2$\,steradians,
where $c$ is the speed of light.  Thus, to cover 3\,deg$^2$ the required
number of pencil beams is $(3\pi/180)^2/(c/\nu D_{\rm core})^2$. Within the
frequency band, the number of pencil beams needs to remain identical as to
enable summing over frequency. In order to have a full coverage of the
3\,deg$^2$, the required number of pencil beams is therefore obtained by
setting $\nu$ to $\nu_{\rm max}$. The number of operations per second to
perform the beam forming of the AA is then given by (see also
Eq.~\ref{eq:beamformingAA})
\begin{equation}
 N_{\rm osb} = F_c N_{\rm dish} N_{\rm pol} B\cdot3\left(\frac{\pi}{180c}\right)^2 D_{\rm core}^2\nu_{\rm max}^2.
\label{eq:beamformingAA2}
\end{equation}

\subsection*{Derivation of figure of merit}
Neglecting numerical factors and physical constants, the figure of merit
expresses the speed of a survey for a given sensitivity of the telescope. It
therefore scales with aperture efficiency ($\eta$), the FoV, the bandwidth
($B$), the square of the collecting area ($A$) and the reciprocal of the
square of the system temperature ($T_{\rm sys}$). For a pulsar survey it also
scales with the square of the flux of pulsars, which scales as $\nu^\alpha$,
with $\nu$ the observation frequency and $\alpha$ the average spectral index
of radio pulsars, which is about --1.6. Thus:
\begin{equation}
M = \eta B {\rm FoV} \left(\frac{A\nu^\alpha}{T_{\rm sys}}\right)^2.
\label{eq:Merit_derivation}
\end{equation}
The FoV is given by $\xi(\lambda D_{\rm dish}^{-1})^2$, where $\xi$ is the factor
by which the phased array feed extends the FoV and $D_{\rm dish}$ is the
diameter of the antennas. This is equal to $\xi (c \nu^{-1} D_{\rm
  dish}^{-1})^2$. The collecting area can be expressed as $A = N_{\rm dish}\pi
D_{\rm dish}^2/4$, where $N_{\rm dish}$ is the number of dishes used in the survey. $T_{\rm sys}$
is the sum of the telescope temperature and the sky temperature, the latter
scaling with frequency as $\nu^{-2.6}$. Putting all this in
Eq.~\ref{eq:Merit_derivation} and removing all constants, yields:
\begin{equation}
 M = \eta \xi B\left(\frac{N_{\rm dish} D_{\rm dish}}{T_{\rm tel}+T_{\rm
 sky}^{400}\left(\frac{\nu}{400\mathrm{\,MHz}}\right)^{-2.6}}\right)^2\nu^{2\alpha-2}
\end{equation}

\subsection*{Glossary}

The following list summarises the meaning of the most important parameters
in this paper.
\begin{description}
\item[$T_{\rm sys}$:] system temperature (including contributions from receiver
  and sky background. 
\item[$T_{\rm sky}^{400}$:] sky temperature at 400 MHz.
\item[FoV:] Field of view, size of the sky area instantaneously covered by the
  telescope.
\item[$\eta$:] aperture efficiency for dishes.
\item[$D_{\rm dish}$:] diameter of a single dish.
\item[$D_{\rm el}$:] distance between the furthest elements that are
  used in an observation. For any pulsar survey in this paper this
  parameter is equal to $D_{\rm core}$.
\item[$D_{\rm core}$:] diameter of the core region.
\item[$N_{\rm osb}$:] number of operations per second required for beam forming.
\item[$N_{\rm dish}$:] total number of dishes.
\item[$F_c$:] fraction of the number of dishes inside the core.
\item[$N_{\rm pol}$:] number of polarisations.
\item[$B$:] observing bandwidth.
\item[$\xi$:] factor by which the phased array feed expands the
  FoV. Note that for a phased array feed this factor is
  frequency dependent.
\item[$\nu_{\rm max}$:] maximum observing frequency.
\item[$a, b$:] parameters appearing in Eq.~\ref{eq:F_c}.
\item[$\cal{D}_{\rm dish}$:] survey data rate for one pencil beam.
\item[$\cal{D}_{\rm AA}$:] total survey data rate for AAs.
\item[$t_{\rm samp}$:] sampling time for pulsar survey.
\item[$\Delta \nu$:] channel bandwidth for pulsar survey.
\item[$N_{\rm bits}$:] number of bits used for digitisation of survey data.
\item[$\nu_{\rm min}$:] lowest frequency used for pulsar survey.
\item[$DM_{\rm max}$:] maximum dispersion measure to be searched or expected.
\item[$N_{\rm DM}$:] number of trial DM-values.
\item[$N_{\rm samp}$:] number of samples in each survey time series.
\item[$N_{\rm acc}$:] number of trial accelerations to be searched.
\item[$N_{\rm pbAA}$:] number of pencil beams for AAs.
\item[$N_{\rm oa}$:] number of operations to search one pencil beam
in acceleration searches.
\item[$M$:] survey-speed figure-of-merit for searches with dishes.
\item[$t_{\rm total}$:] total observing time for a survey.
\item[$t_{\rm point}$:] observing time per survey pointing.
\item[$\Omega_{\rm sur}$:] total sky (solid angle) for survey.
\item[$\Omega_{\rm FoV}$:] total FoV for dishes.
\end{description}

\bibliographystyle{aa} 
\bibliography{ska}

\end{document}